\documentclass[12pt]{iopart}

\usepackage{graphicx}

\begin{document}

\title[Optimal two-mode attacks for two-way CV-QKD protocols]{Numerical simulation of the optimal two-mode attacks for two-way continuous-variable quantum cryptography in reverse reconciliation}

\author{Yichen Zhang$^{1}$ \& Zhengyu Li$^{2}$ \& Yijia Zhao$^{1}$ \& Song Yu$^{1}$ \& Hong Guo$^{2}$}

\address{$^{1}$ State Key Laboratory of Information Photonics and Optical Communications, Beijing University of Posts and Telecommunications, Beijing 100876, China\\
$^2$ State Key Laboratory of Advanced Optical Communication Systems and Networks, School of Electronics Engineering and Computer Science, Center for Quantum Information Technology, Peking University, Beijing 100871, China}
\ead{yusong@bupt.edu.cn}
\vspace{10pt}
\begin{indented}
\item[]\today
\end{indented}

\begin{abstract}
We analyze the security of the two-way continuous-variable quantum key distribution protocol in reverse reconciliation against general two-mode attacks, which represent all accessible attacks at fixed channel parameters. Rather than against one specific attack model, the expression of secret key rates of the two-way protocol are derived against all accessible attack models. It is found that there is an optimal two-mode attack to minimize the performance of the protocol in terms of both secret key rates and maximal transmission distances. We identify the optimal two-mode attack, give the specific attack model of the optimal two-mode attack and show the performance of the two-way protocol against the optimal two-mode attack. Even under the optimal two-mode attack, the performances of two-way protocol are still better than the corresponding one-way protocol, which shows the advantage of making a double use of the quantum channel and the potential of long-distance secure communication using two-way protocol.

\end{abstract}

%
%
%
%
%

\section{Introduction}
Quantum key distribution (QKD)~\cite{Gisin_RevModPhys_2002,Scarani_RevModPhys_2009} is one of the most practical applications in the field of quantum information. Its goal is to establish a secure key between two legitimate partners, usually named Alice and Bob. Continuous-variable quantum key distribution (CV-QKD)~\cite{Braunstein_RevModPhys_2005,Xiang-Bin_PhysReport_2007,Weedbrook_RevModPhys2012} has attracted much attention in the past few years~\cite{Weedbrook_RevModPhys2012,Jouguet_nature_2013,Zhengyu_PhysRevA_2014,My_PhysRevA_2014,Pirandola_NatPhoton_2015} mainly because it only uses standard telecom components. A CV-QKD protocol based on Gaussian-modulated coherent states~\cite{Grosshans_PhysRevLett_2002,Weedbrook_PhysRevLett_2004} has been proved to be secure against arbitrary general attacks~\cite{Navascues_PhysRevLett_2006,Garcia_PhysRevLett_2006,Pirandola_PhysRevLett_2009,Renner_PhysRevLett_2009,Leverrier_PhysRevLett_2013,Leverrier_PhysRevLett_2015} and experimentally demonstrated~\cite{Jouguet_nature_2013,grosshans_nature_2003,Lance_PRL_2005,Lodewyck_PhysRevA_2007,Khan_PhysRevA_2013}.

To enhance the tolerable excess noise of CV-QKD, compared to the typical one-way schemes, the two-way CV-QKD protocol was proposed~\cite{pirandola_nature_2008}. Afterward, a more feasible two-way CV-QKD protocol was proposed by replacing Alice's displacement operation with a beam splitter, which leads to a protocol that is easier to analyze when considering channel estimation~\cite{sunmaozhu_WorldScientific_2012}. In standard CV-QKD protocols the quantum communication is one way, i.e., quantum systems are sent from Alice to Bob. While in two-way protocols, this process is bidirectional, with the systems transformed by Alice and sent back to Bob. The use of two-way quantum communication can increase the secure key rate, transmission distance, and the robustness to noise~\cite{pirandola_nature_2008,sunmaozhu_WorldScientific_2012}. As a result, bosonic channels which are too noisy for one-way protocols may become secure for two-way protocols~\cite{Weedbrook_RevModPhys2012}.

However, when comparing the performance of two-way CV-QKD protocol with one-way protocol in numerical simulations, all of the works assumed the eavesdropper performs two independent attacks~\cite{pirandola_nature_2008,sunmaozhu_WorldScientific_2012,Weedbrook_PhysRevA_2014} or give a specific attack model~\cite{My_JPB_2014,sunmaozhu_JPB_2013}. It makes us to wander does the outperformance of two-way protocol against a specific attack mean two-way protocol really have advantages than one-way protocol? Do we make a fair comparison? Or it is just because the chosen attack model is powerless?

Recently, it is very interesting to see that the security of the original two-way CV-QKD protocol against two-mode attacks has been studied assuming direct
reconciliation~\cite{Carlo_PhysRevA_2015}, and the general immunity and superadditivity of the original protocol has also been studied~\cite{Carlo_SciRep_2016}. Inspired by the method of~\cite{Carlo_PhysRevA_2015}, in this paper, we analyze the security of the modified two-way CV-QKD protocol in reverse reconciliation against general two-mode attacks, including two independent attacks, all separable attacks and all entangled attacks. Normally, reverse reconciliation is more useful than direct reconciliation in practice~\cite{Weedbrook_RevModPhys2012}. Against all accessible two-mode attacks, the expression of secret key rates of two-way CV-QKD protocol using coherent states are derived under reverse reconciliation. Then we evaluate and compare the performance of the two-way protocol against different attacks and identify the optimal attack model, resulting the lowest secret key rate, for the eavesdropper at different transmission distance. We also show the performance of the two-way CV-QKD protocol against the optimal two-mode attack. Finally, the performances of the two-way CV-QKD protocol against the optimal attack are compared with the performances of the one-way version of the scheme and show that the two-way CV-QKD protocol still achieves higher secret key rate than one-way protocol. Thus, the two-way protocol is still able to distribute secret keys in communication lines which are too noisy for the corresponding one-way protocol.

The paper is organized as follows. In Sec.~\ref{sec:2}, we review the basic notions of the entanglement-based scheme of the two-way CV-QKD protocol and the general two-mode attacks and identify different type of attacks. In Sec.~\ref{sec:3}, we derive the expression of secret key rates of the coherent-state based two-way CV-QKD protocol. In Sec.~\ref{sec:4}, the simulation results against different attacks are provided and the performances of the two-way protocol against the optimal attack are compared with one-way protocol to show the advantage of two-way scheme. Our conclusions are drawn in Sec.~\ref{sec:5}.

\begin{figure*}[t]
\centerline{\includegraphics[width=16.0cm]{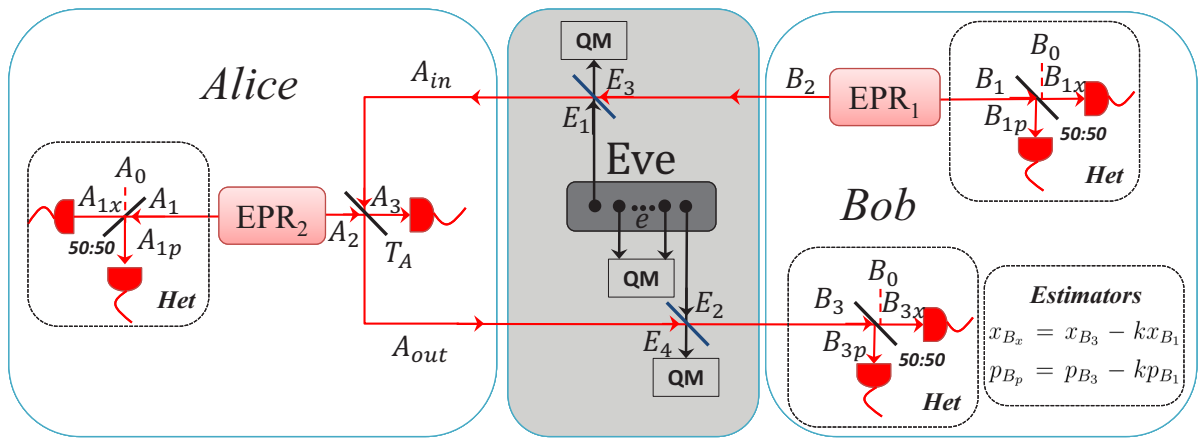}}
\caption{ (Color online) The entanglement-based scheme of two-way CV-QKD protocol using Gaussian-modulated coherent states against two-mode attacks, where the quantum channel is fully controlled by Eve. However, Eve has no access to the apparatuses in Alice's and Bob's stations.}\label{fig1}
\end{figure*}

\section{\label{sec:2} Entanglement-based model of two-way CV-QKD protocols against two-mode attacks}

In this section, we first present basic notions of the entanglement-based scheme of the two-way CV-QKD protocol using coherent states. Then we describe the general two-mode attacks and identify different type of attacks.

\subsection{Entanglement-based model of two-way CV-QKD protocols}

The entanglement-based scheme of the two-way CV-QKD protocol using coherent states is illustrated in~Fig.~\ref{fig1} and can be described as follows:
\\
\\{Step 1:} Bob initially prepares an EPR pair (EPR1 with variance $V_{B}$, where the shot noise variance is normalized to $1$), keeps the mode $B_{1}$ and sends the other mode $B_{2}$ to Alice through the channel where Eve may perform her attack.
\\
\\{Step 2:} Alice prepares another EPR pair (EPR2 with variance $V_{A}$). She keeps the mode $A_{1}$ and measures it using heterodyne detection to get the variables $x_{A_{1}}$ and $p_{A_{1}}$. She then couples mode $A_{2}$ and the received mode $A_{in}$ from Bob with a beam splitter (transmittance: $T_{A} \in [0,1]$). Alice then sends mode $A_{out}$ back to Bob where Eve may perform her attack again. Alice measures another mode $A_{3}$ with homodyne detection for parameter estimation~\cite{sunmaozhu_WorldScientific_2012}.
\\
\\{Step 3:} Bob measures his original mode $B_{1}$ using heterodyne detection to get the variables $x_{B_{1}}$ and $p_{B_{1}}$. He also measures the received mode $B_{3}$ with heterodyne detection to get $x_{B_{3}}$ and $p_{B_{3}}$.
\\
\\{Step 4:} Bob uses ${x_{B_x}} = {x_{B_3}} - {k} x_{B_{1}}$  and ${p_{B_p}} = {p_{B_3}} - {k} p_{B_{1}}$ to construct the estimator to Alice's corresponding variable $x_{{A_1}}$ and $p_{{A_1}}$, where ${k}$ is the parameter used to optimize Bob's estimator of Alice¡¯s corresponding value. Then Alice and Bob proceed with classical data postprocessing including reconciliation and privacy amplification. Here we use reverse reconciliation~\cite{grosshans_nature_2003}.
\\

The most general eavesdropping strategy of two-way CV-QKD protocol is coherent attacks, which involve a unitary applied to all modes over all uses of the protocol. However, this could be reduced to collective attacks by assuming that Alice and Bob perform random permutations on their data~\cite{Renner_PhysRevLett_2009}. When the eavesdropper perform collective attacks, she interacts independently and identically with each quantum signal over every uses. In the entanglement-based representation of two-way CV-QKD protocol, this means that the joint state ${\rho _{{A^n}{B^n}}}$ has an identical and independently distributed (i.i.d.) structure ${\rho _{{A^n}{B^n}}} = \rho _{AB}^{ \otimes n}$, where n amounts the number of the uses of the protocol.

What's more, the most general collective attack against two-way CV-QKD protocol is a joint attack involving two channels, the forward and backward channels. In each use of the protocol, Eve could intercept the two modes, one is the output of Bob side in the forward channel (mode $B_2$ in Fig.~\ref{fig1}) and the other is the output of Alice side in the backward channel (mode $A_{out}$ in Fig.~\ref{fig1}), and make them intercept with an ensemble of ancillary vacuum modes via a general unitary $U$. The remaining modes are stored in a quantum memory which will be measured at the end of the protocol.

The description of this attack can be further simplified. Since the protocol is based on the Gaussian modulation, its optimal eavesdropping attack is based on a Gaussian unitary $U$. Thus, the security of the protocol can be reduced to studying a two-mode Gaussian attack against two channels, which is depicted in Fig.~\ref{fig1}.

\subsection{Two-mode attack strategy}

In this two-mode Gaussian attack, the two output modes, $B_2$ and $A_{out}$, are mixed with two ancillary modes, $E_1$ and $E_2$, by two beam splitters with transmissivities $T_1$ and $T_2$, respectively. These ancillary modes belong to a reservoir of ancillas ($E_1$, $E_2$ and an extra set $e$) which is globally described by a pure Gaussian state. The reduced state $\rho_{E_1 E_2}$ of the injected ancillas is a correlated thermal state with zero mean and covariance matrix in the normal form

\begin{equation}\label{eq1}
{\gamma_{{E_1}{E_2}}} = \left( {\begin{array}{*{20}{c}}
   {{V_{{E_1}}} \cdot {{\rm{I}}_2}} & {{C_{{E_1}{E_2}}}}  \\
   {{C_{{E_1}{E_2}}}} & {{V_{{E_2}}} \cdot {{\rm{I}}_2}}  \\
\end{array}} \right),
\end{equation}
where ${V_{E_1}}$ and ${V_{E_2}}$ are the variances of the thermal noise affecting each channel, ${C_{{E_1}{E_2}}} =  diag \left( C_x, C_p \right)$ is the correlation parameters between two ancillas. The various parameters ${V_{E_1}}$, ${V_{E_2}}$, $C_x$ and $C_p$ must satisfy the physical constraints~\cite{Pirandola_NatPhoton_2015, Pirandola_PhysRevA_2009, Pirandola_OpenSystInfDyn_2013, Ottaviani_PhysRevA_2015}:
\begin{equation}\label{eq2}
{\gamma_{{E_1}{E_2}}} > 0,\;\;{\nu _ - } \ge 1,
\end{equation}
where the positivity ${\gamma_{{E_1}{E_2}}} > 0$ is equivalent to the positivity of the principal minors of the matrix of Eq.~\ref{eq1}, ${\nu _ - } = \sqrt {0.5\left( {\Delta \left( {{\gamma_{{E_1}{E_2}}}} \right) - \sqrt {\Delta {{\left( {{\gamma_{{E_1}{E_2}}}} \right)}^2} - 4\det {\gamma_{{E_1}{E_2}}}} } \right)} $, and $\Delta \left( {{\gamma_{{E_1}{E_2}}}} \right): = V_{{E_1}}^2 + V_{{E_2}}^2 + 2{C_x}{C_p}$.

\begin{figure*}[t]
\centerline{\includegraphics[width=9cm]{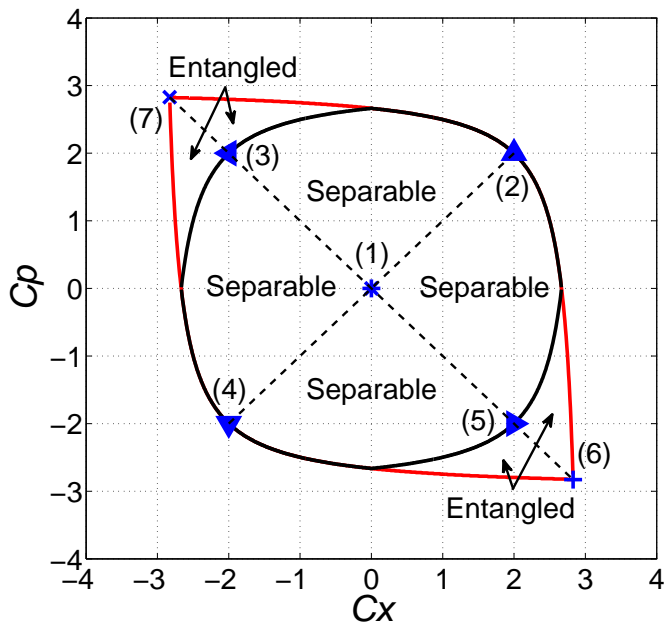}}
\caption{ (Color online) Correlation plane for different types of attacks under fixed variances ${V_{E_1}} = {V_{E_2}} = 3$, where the inner are corresponds to separable attacks, while the two peripheral areas correspond to entangled attacks. The number points represent the specific attacks: point (1) represents \emph{Independent Attack}; point (2), (3), (4) and (5) represent \emph{Separable Attack} ($C_{sep}^{\max } = 2$); point (6) and (7) represent \emph{Entangled Attack} ($C_{ent}^{\max } = \sqrt{8} $) . }\label{fig2}
\end{figure*}

In the two-way CV-QKD protocols, normally the excess noises of two channels are fixed, which means the variances, ${V_{E_1}}$ and ${V_{E_2}}$, are fixed for every transmissivity $T_1$ and $T_2$. Thus, the remaining degrees of freedom in the two-mode Gaussian attack are the correlation parameters $C_x$ and $C_p$, which can be represented as a point on a correlation plane. Each point of this plane describes an attack. Among all these accessible attacks, those satisfying the further condition
\begin{equation}\label{eq3}
{{\tilde \nu }_ - } \ge 1,
\end{equation}
are separable attacks (${\gamma_{{E_1}{E_2}}}$ seperable), while those violating the condition of Eq.~\ref{eq3} are entangled attacks (${\gamma_{{E_1}{E_2}}}$ entangled), where ${{\tilde \nu }_ - } = \sqrt {0.5\left( {\tilde \Delta \left( {{\gamma _{{E_1}{E_2}}}} \right) - \sqrt {\tilde \Delta {{\left( {{\gamma _{{E_1}{E_2}}}} \right)}^2} - 4\det {\gamma _{{E_1}{E_2}}}} } \right)} $, and $\tilde \Delta \left( {{\gamma _{{E_1}{E_2}}}} \right): = V_{{E_1}}^2 + V_{{E_2}}^2 - 2{C_x}{C_p}$. Fig.~\ref{fig2} are a numerical representation of the correlation plane under fixed variances: ${V_{E_1}} = 3$ and ${V_{E_2}} = 3$. Then we can identify the following attacks: Separable Attack, Independent Attack and Entangled Attack, which is detailed describe in Ref.~\cite{Pirandola_NatPhoton_2015, Ottaviani_PhysRevA_2015}. Here we just give the some basic knowledge which will be used in the following analysis. As illustrated in Fig.~\ref{fig2}, the point (1) represents the independent attack; the point (2), (3), (4) and (5) represent four specific separable attacks; the point (6) and (7) represent two specific entangled attacks.

\subsection{Security analysis of two-way CV-QKD protocol against two-mode attacks}

In this subsection, we will derive the secure bound of the two-way protocol using coherent states against two-mode Gaussian attacks.

From the information-theoretic perspective, the asymptotic secret key rate ${K}$ against two-mode Gaussian attacks in reverse reconciliation is given by~\cite{Devetak_ProcRSoc_2005}
\begin{equation}
 {K} = \beta I\left( {A:B} \right) - \chi \left( {B:E} \right),
\end{equation}
where $\beta$ is the reconciliation efficiency, $I(A:B)$ is the classical mutual information between Alice and Bob, $\chi(B:E)$ is the the Holevo bound between Eve and Bob~\cite{Nielsen_QCQI}
\begin{equation}\label{eq5}
\chi \left( {B:E} \right) = S\left( {{\rho _E}} \right) - \sum\nolimits_{{m_B}} {p\left( {{m_B}} \right)S\left( {\rho _E^{{m_B}}} \right)} ,
\end{equation}
where $S(\rho)$ is the von Neumann entropy of the quantum state $\rho$, $m_B$ is Bob's measurement result, and it can take the form $m_B = x_B$ for homodyne detection or the form $m_B = x_B, p_B$ for heterodyne detection. $p\left( m_B \right)$ is the probability density of the measurement result, $\rho _{E}^ {m_B}$ is the corresponding state of Eve's ancillary conditioned on Bob's measurement result.

The overall state can be described by the covariance matrix, which is defined by
\begin{equation}
{\gamma _{ij}} = {\rm{Tr}}\left[ {\hat \rho \left\{ {\left( {{{\hat r}_i} - {d_i}} \right),\left( {{{\hat r}_j} - {d_j}} \right)} \right\}} \right] ,
\end{equation}
where ${{\hat r}_{2i-1}} = {{\hat x}_i}$, ${{\hat r}_{2i}} = {{\hat p}_i}$, ${d_i} = \left\langle {{{\hat r}_i}} \right\rangle  = {\rm{Tr}}\left[ {\hat \rho {{\hat r}_i}} \right]$, $\hat \rho$ is the density matrix, and ${}$ denotes the anticommutator. Before channel transmission, the covariance matrix ${\gamma _{{B_1}{A_1}{A_2}{B_2}}}$ is
\begin{equation}
\left[ {\begin{array}{*{20}{c}}
   {{V_B} \cdot {\rm{I}}} & {0 \cdot {\rm{I}}} & {0 \cdot {\rm{I}}} & {\sqrt {\left( {V_B^2 - 1} \right)}  \cdot {\sigma _z}}  \\
   {0 \cdot {\rm{I}}} & {{V_A} \cdot {\rm{I}}} & {\sqrt {\left( {V_A^2 - 1} \right)}  \cdot {\sigma _z}} & {0 \cdot {\rm{I}}}  \\
   {0 \cdot {\rm{I}}} & {\sqrt {\left( {V_A^2 - 1} \right)}  \cdot {\sigma _z}} & {{V_A} \cdot {\rm{I}}} & {0 \cdot {\rm{I}}}  \\
   {\sqrt {\left( {V_B^2 - 1} \right)}  \cdot {\sigma _z}} & {0 \cdot {\rm{I}}} & {0 \cdot {\rm{I}}} & {{V_B} \cdot {\rm{I}}}  \\
\end{array}} \right]
\end{equation}
where $V_B$ is the variance of EPR1 and $V_A$ is the variance of EPR2. The channel transmission relationship are as follow,
\begin{equation}
\left\{ \begin{array}{l}
 {r_{{A_{in}}}} = \sqrt {{T_1}} {r_{{B_2}}} + \sqrt {1 - {T_1}} {r_{{E_1}}} \\
 {r_{{A_3}}} =  - \sqrt {1 - \eta } {r_{{A_{in}}}} + \sqrt \eta  {r_{{A_2}}} \\
 {r_{{A_{out}}}} = \sqrt \eta  {r_{{A_{in}}}} + \sqrt {1 - \eta } {r_{{A_2}}} \\
 {r_{{B_3}}} = \sqrt {{T_2}} {r_{{A_{out}}}} + \sqrt {1 - {T_2}} {r_{{E_2}}} \\
 \end{array} \right.
\end{equation}

After two-mode Gaussian attacks, the covariance matrix ${\gamma _{{B_1}{A_1}{A_2}{B_2}}}$ is changed into  ${\gamma _{{B_1}{A_1}{A_3}{B_3}}}$, which is given by

\begin{equation}\label{eq}
\left[ {\begin{array}{*{20}{c}}
   {{V_B}{\rm{I}}} & 0 & { - T'{C_B}{\sigma _z}} & {\sqrt \eta  T{C_B}{\sigma _z}}  \\
   0 & {{V_A}{\rm{I}}} & {\sqrt \eta  {C_A}{\sigma _z}} & {T'{C_A}{\sigma _z}}  \\
   { - T'{C_B}{\sigma _z}} & {\sqrt \eta  {C_A}{\sigma _z}} & {{V_{{A_3}}}} & {{C_{{A_3}{B_3}}}}  \\
   {\sqrt \eta  T{C_B}{\sigma _z}} & {T'{C_A}{\sigma _z}} & {{C_{{A_3}{B_3}}}} & {{V_{{B_3}}}}  \\
\end{array}} \right]
\end{equation}
where the forward and backward channel are assumed to have identical transmissitivity $T_1 = T_2 =T$ and same excess noise $V_{E_1} = V_{E_2} = V_E$. The parameters $T' = \sqrt {T\left( {1 - \eta } \right)}$, ${C_A} = \sqrt {\left( {V_A^2 - 1} \right)}$, ${C_B} = \sqrt {\left( {V_B^2 - 1} \right)}$. The matrices ${V_{{A_3}}} = \left[ {\eta {V_A} + \left( {1 - \eta } \right)\left( {T{V_B} + \left( {1 - T} \right){V_E}} \right)} \right]{\rm{I}}$, ${V_{{B_3}}} =  diag \left( {V_{{B_3}}^x}, {V_{{B_3}}^p}\right) $ and ${C_{{A_3}{B_3}}} =  diag \left( {C_{{A_3}{B_3}}^x}, {C_{{A_3}{B_3}}^p}\right) $, which is given by

\begin{equation}
\left\{ {\begin{array}{*{20}{c}}
   {V_{{B_3}}^x = V' + 2{C_x}\left( {1 - T} \right)\sqrt {T\eta } }  \\
   {V_{{B_3}}^p = V' + 2{C_p}\left( {1 - T} \right)\sqrt {T\eta } }  \\
   {C_{{A_3}{B_3}}^x = C' - {C_x}\left( {1 - T} \right)\sqrt {1 - \eta } }  \\
   {C_{{A_3}{B_3}}^p = C' - {C_p}\left( {1 - T} \right)\sqrt {1 - \eta } }  \\
\end{array}} \right.
\end{equation}
where $V' = T\left( {1 - \eta } \right){V_A} + {T^2}\eta {V_B} + \left\{ {1 - T\left[ {1 - \eta \left( {1 - T} \right)} \right]} \right\}{V_E}$ and $C' = \sqrt {T\eta \left( {1 - \eta } \right)} {V_A} - T\sqrt {T\eta \left( {1 - \eta } \right)} {V_B} - \left( {1 - T} \right)\sqrt {T\eta \left( {1 - \eta } \right)} {V_E}$

For two-way CV-QKD protocol using coherent states, Alice measures mode $A_1$ to get the variables {$x_{A_{1}}$, $p_{A_{1}}$} using heterodyne detectors. While Bob measures modes $B_1$ and $B_3$ to get the variables {$x_{B_{1}}$, $p_{B_{1}}$} and {$x_{B_{3}}$, $p_{B_{3}}$} using heterodyne detectors. Then he uses the estimators {${x_{B_x}} = {x_{B_{3}}} - {k} x_{B_{1}}$, ${p_{B_p}} = {p_{B_{3}}} - {k} p_{B_{1}}$} to construct {$x_{{A_{1}}}$, $p_{{A_{1}}}$} at the same time.

The classical mutual information between Alice and Bob becomes
\begin{equation}
\begin{array}{l}
 I\left( {A:B} \right) = {I_x}\left( {A:B} \right) + {I_p}\left( {A:B} \right) \\
 \quad \quad \quad \;{\rm{ = }}\frac{1}{2}\log \frac{{{V_{{A_{1x}}}}}}{{{V_{{A_{1x}}{\rm{|}}{B_x}}}}} + \frac{1}{2}\log \frac{{{V_{{A_{1p}}}}}}{{{V_{{A_{1p}}{\rm{|}}{B_p}}}}} \\
 \end{array},
\end{equation}
where $V_{A_{1x}} = V_{A_{1p}} = \frac{1}{2} \left( V_{A_1} + 1 \right)$, $V_{A_{1x} | x_{B_x}} = \frac{1}{2} \left( V_{A_{1} | x_{B_x}} + 1 \right)$ and $V_{A_{1p} | p_{B_p}} = \frac{1}{2} \left( V_{A_{1} | p_{B_p}} + 1 \right)$.

The calculation of $\chi \left( {B:E} \right)$ is more complex. For heterodyne detection, we have $\chi_x \left( {B:E} \right) = S\left( E \right) -S\left( E| x_{B_x}, p_{B_p} \right)$. Assuming Eve is able to purify Alice and Bob's system, we have $ S\left( E \right) = S\left( B_1 A_1 A_3 B_3 \right)$ and $S\left( E| x_{B_x} , p_{B_p} \right) = S\left( B_1 A_1 A_3 B_3 | x_{B_x} , p_{B_p} \right)$. The $S\left( B_1 A_1 A_3 B_3 \right)$ and $S\left( B_1 A_1 A_3 B_3 | x_{B_x} , p_{B_p} \right)$ can be calculated by the symplectic eigenvalues of the covariance matrices $\gamma_{ B_1 A_1 A_3 B_3 }$ and $\gamma_{ B_1 A_1 A_3 B_3 | x_{B_x} , p_{B_p}}$, which is derived from the following method.

Given an arbitrary $N$-mode covariance matrix $\gamma$, there exists a symplectic matrix $S$ such that
\begin{equation}
\gamma  = S{\gamma ^ \oplus }{S^T},\quad {\gamma ^ \oplus } = \mathop  \oplus \limits_{k = 1}^N {\lambda _k} \cdot {{\rm I}_2},
\end{equation}
where the diagonal matrix ${\gamma ^ \oplus }$ is called the Williamson form of $\gamma$, and the $N$ positive quantities $\lambda _k$ are called the symplectic eigenvalues of $\gamma$ \cite{Weedbrook_RevModPhys2012}. Here the symplectic spectrum $\left\{ {{\lambda _k}} \right\}_{k = 1}^N$ can be easily computed as the standard eigenspectrum of the matrix $\left| {i\Omega \gamma } \right|$ \cite{Weedbrook_RevModPhys2012}, where the modulus must be understood in the operational sense. Here $\Omega$ is the symplectic form $\Omega  = \mathop  \oplus \limits_{k = 1}^N {\kern 1pt} {\kern 1pt} \left[ {\begin{array}{*{20}{c}}
   0 & 1  \\
   { - 1} & 0  \\
\end{array}} \right].$

The covariance matrix $\gamma_{ B_1 A_1 A_3 B_3}$ is given by Eq.~\ref{eq}, while the derivation of $\gamma_{ B_1 A_1 A_3 B_3 | x_{B_x} , p_{B_p}}$ need to concern Bob's post-processing strategy. In coherent-state based protocol, Bob uses ${x_{B_x}} = {x_{B_3}} - {k} x_{B_{1}}$ and ${p_{B_p}} = {p_{B_{3}}} - {k} p_{B_{1}}$ to construct $x_{{A_{1}}}$ and $p_{{A_{1}}}$ at the same time, where ${k} = \sqrt {0.5 {T^2}\eta \frac{{{V_B} - 1}}{{{V_B} + 1}}} $. Thus, we are not able to derive $\gamma_{ B_1 A_1 A_3 B_3 | x_{B_x}}$ directly. As developed in~\cite{My_JPB_2014}, we transform modes $B_{3x}$ and $B_{1x}$ into modes $B_x$ and $B_4$ with a CNOT gate $\Gamma_x $, and transform modes $B_{3p}$ and $B_{1p}$ into modes $B_p$ and $B_5$ with a CNOT gate $\Gamma_p $. We should emphasize that this is a virtual operation in the entanglement-based scheme, while in its corresponding Prepare \& Measurement scheme Bob does not perform such operations.

Since we have the covariance matrices $\gamma_{ B_1 A_1 A_3 B_3 }$ and $\gamma_{ B_1 A_1 A_3 B_3 | x_{B_x}, p_{B_p}}$, we could calculate the symplectic eigenvalues of them by the method mentioned before. Thus, the expression for Eq.~\ref{eq5} can be further simplified as follows:

\begin{equation}
\chi \left( {B:E} \right)  = \sum\limits_{i = 1}^4 {G\Big(\frac{{{\lambda _i} - 1}}{2}\Big)}  - \sum\limits_{i = 5}^{8} {G\Big(\frac{{{\lambda _i} - 1}}{2}\Big)},
\end{equation}
where $G(x) = (x + 1)\log_2 (x + 1) - x\log_2 x$, ${\lambda _{1 - 4}}$ are the symplectic eigenvalues of the covariance matrix $\gamma_{ B_1 A_1 A_3 B_3 }$ and ${\lambda _{5 - 8}}$ are the symplectic eigenvalues of the covariance matrix $\gamma_{ B_1 A_1 A_3 B_3 | x_{B_x}, p_{B_p}}$.

\section{\label{sec:3} Optimal two-mode attack strategy}

In this section, to find the optimal two-mode attack strategy against two-way CV-QKD protocols, the performance of two-way CV-QKD protocols are compared under different types two-mode attack strategy.

\begin{figure*}[!t]
\centerline{\includegraphics[width=16cm]{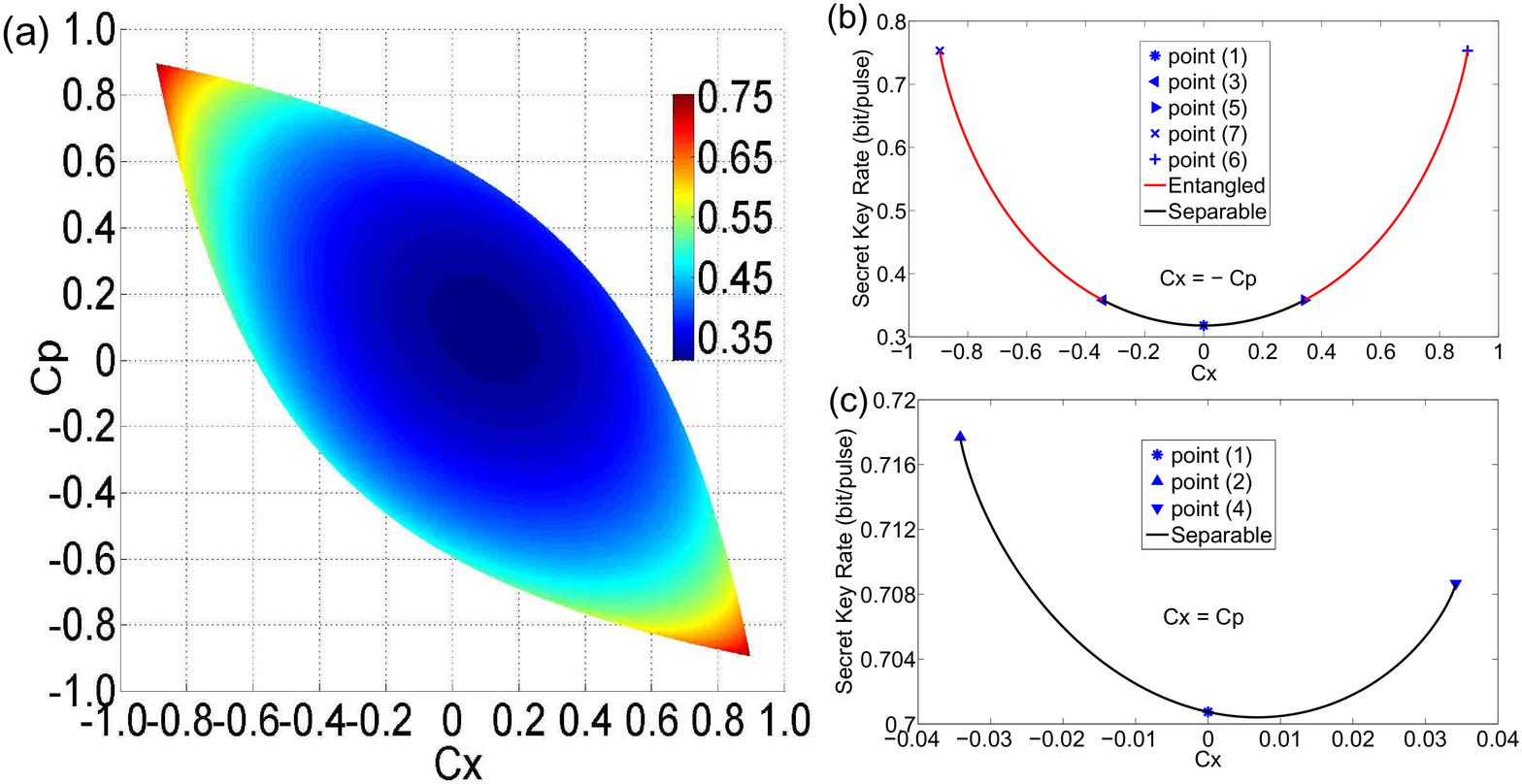}}
\caption{(Color online) (a) Secret key rates of two-way CV-QKD protocol using coherent states against all accessible attacks under $10 km$ distance, where different colors correspond to different values of the rate. The secret key rate is symmetric with respect to the bisector $C_x = C_p$. (b) Specific case of the left figure where $C_x = C_p$. (c) Specific case of the left figure where $C_x = - C_p$. Here we use the reconciliation efficiency $\beta = 0.95$~\cite{Jouguet_nature_2013}, modulation variance $V_A = V_B = 20$, $\varepsilon = 0.2$ and $\eta = 0.75$. Point (1) represents \emph{Independent Attack}; point (2), (3), (4) and (5) represent \emph{Separable Attack}; point (6) and (7) represent \emph{Entangled Attack}. The various attacks (1) - (7) are classified in Sec. \ref{sec:2} and displayed in Fig. \ref{fig2}.
}\label{fig3}
\end{figure*}

We first analyze the secret key rates of the protocol against all accessible attacks at fixed transmission distance $d = 10 km, 20 km, 30 km$, where we only put the result at $d = 10 km$ here and put the results at $d = 20 km, 30 km$ in the appendix. The parameters affecting the value of the secret key rate are the reconciliation efficiency $\beta$, the variance of Alice's and Bob's modulation: $(V_A-1)$ and $(V_B-1)$, the transmittance of the beam splitter at Alice's side $\eta $ and the transmission efficiency $T$. The parameters ${V_A}$, ${V_B}$, $\beta$, $\eta $ and ${\upsilon _{el}}$ are fixed in all simulations. Here, we choose the variance $V_A = V_B = 20$, $\beta = 1$, $\eta = 0.75$ as the value of the beam splitter transmittance at Alice's side and channel noise $\varepsilon = 0.2$. As illustrated in Fig.~\ref{fig3} (a), different colors correspond to different values of the secret key rate, where color red corresponds to higher value of the rate, while color blue corresponds to lower value of the rate. It is found that the secret key rate is symmetric with respect to the bisector $C_x = C_p$, which is coincident with the security analysis in Sec.2.3. What's more, in order to be more clear, we plot the specific cases where $C_x = - C_p$ and $C_x = C_p$ in Fig.~\ref{fig3} (b) and (c), respectively. The minimum key rate associated with the optimal attack correspond to the two-mode symmetric separable attack with $C_x = C_p = C_{opt}^{10km} = 0.0078$ when we fix the channel excess noise $\varepsilon = 0.2$. The situation at distance $d = 20 km, 30 km$ is illustrated in the appendix. The optimal attack is the two-mode symmetric separable attack $C_x = C_p = C_{opt}^{20km} = 0.0073$ at $d = 20 km$, and is the two-mode symmetric separable attack with $C_x = C_p = C_{opt}^{30km} = 0.0039$ at $d = 30 km$

\begin{figure*}[t]
\centerline{\includegraphics[width=12cm]{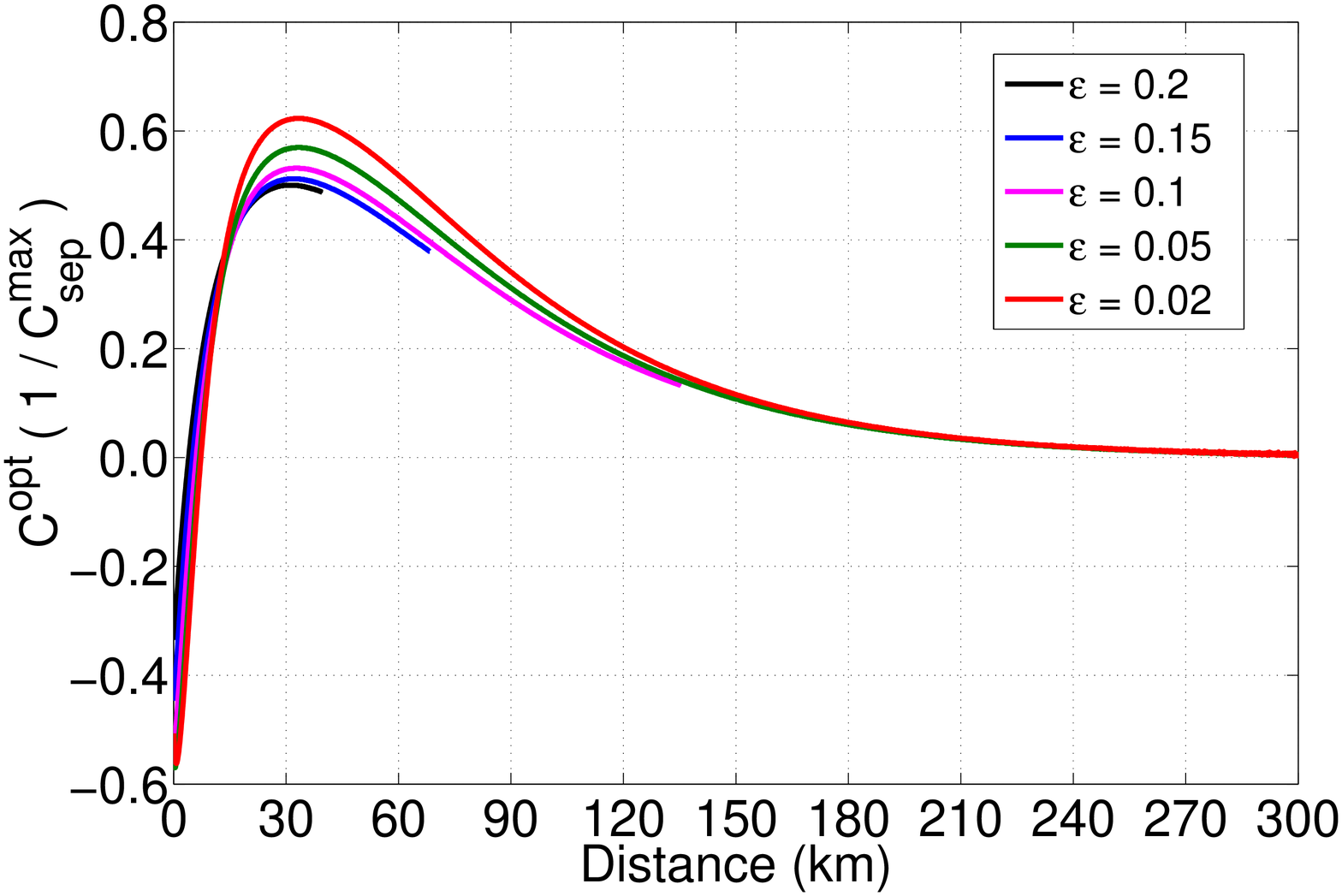}}
\caption{(Color online) The normalized correlation parameters of the optimal two-mode attacks as a function of the distance under different channel excess noise $\varepsilon = 0.2, 0.15, 0.1, 0.05, 0.02$. Here we use the modulation variance $V = V_A = V_B = 20$ and $\eta = 0.75$.
}\label{fig4}
\end{figure*}

\begin{figure*}[b]
\centerline{\includegraphics[width=13cm]{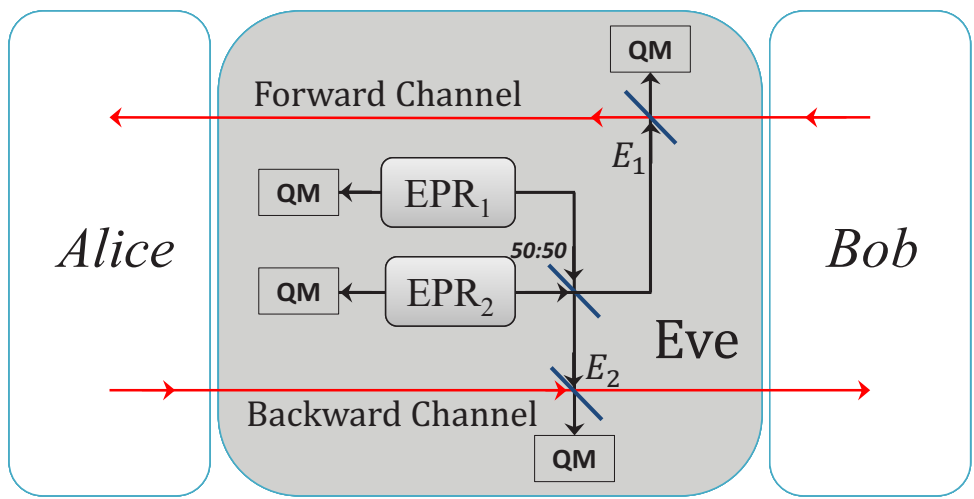}}
\caption{(Color online) The specific two-mode attack model of the optimal two-mode attack strategy corresponding to Fig.~\ref{fig3}, Fig.~\ref{fig4} and Eq.~\ref{eq16}, where Eve prepares two EPR pairs to generate two ancillary injected modes into the forward and backward channel.
}\label{fig5}
\end{figure*}

Depends on the results of each distance, we plot the normalized correlation parameters of the optimal two-mode attack as a function of the distance under different channel excess noise $\varepsilon = 0.2, 0.15, 0.1, 0.05, 0.02$ in Fig.~\ref{fig4}. The optimal two-mode attack always occurs at the case where the correlation parameter $C_x$ is equal to $C_p$. The normalized correlation parameters ${{C'}_x}$ is defined as ${{C'}_x} = {{{C_x}} \mathord{\left/ {\vphantom {{{C_x}} {C_{sep}^{\max }}}} \right. \kern-\nulldelimiterspace} {C_{sep}^{\max }}}$, where ${{C'}_x} = -1, 0, 1$ correspond the case of points (5), (1), (4) in Fig.~\ref{fig2}. Thus, depending on Eq.~\ref{eq1}, the covariance matrix of the injected ancillas of the optimal two-mode attack becomes

\begin{equation}\label{eq16}
{\gamma _{{E_1}{E_2}}} = \left( {\begin{array}{*{20}{c}}
   {{V_E}\cdot{{\rm{I}}_2}} & {{C_{opt}}\cdot{{\rm{I}}_2}}  \\
   {{C_{opt}}\cdot{{\rm{I}}_2}} & {{V_{{E}}}\cdot{{\rm{I}}_2}}  \\
\end{array}} \right),
\end{equation}
where ${V_E} = 1 + \frac{{T\varepsilon }}{{1 - T}}$.

This covariance matrix represents a specific two-mode attack model, which is illustrated in Fig.~\ref{fig5}. Eve initially prepares two EPR pairs (EPR1 with variance $V_{1} = V_E + C_{opt}$ and EPR2 with variance $V_{2} = V_E - C_{opt}$), keeps one mode of each EPR pairs and then she couples the other mode of the two EPR pairs with a $50: 50$ beam splitter. The output modes of the beam splitter, mode $E_1$ and $E_2$, are two ancillary injected modes of two-mode attack. Finally, Eve mixes modes $E_1$ and $E_2$ with the modes in the forward and backward channels, by two beam splitters with transmissivity $T$, respectively. The remaining modes are stored in the quantum memory which will be measured at the end of the protocol. We note that, no matter what specific model of Eve is, the optimal two-mode attack has the feature that $V_{E_1} = V_{E_2}$, $C_x = C_p$, which is symmetric for x-quadrature and p-quadrature.

\section{\label{sec:4} Outperfomance of two-way CV-QKD protocols}

In the following, we compare the two-way CV-QKD protocol with their one-way counterpart to examine whether two-way protocol remains advantageous when the optimal two-mode attack strategy is used against the two-way protocol.

\begin{figure*}[t]
\centerline{\includegraphics[width=16cm]{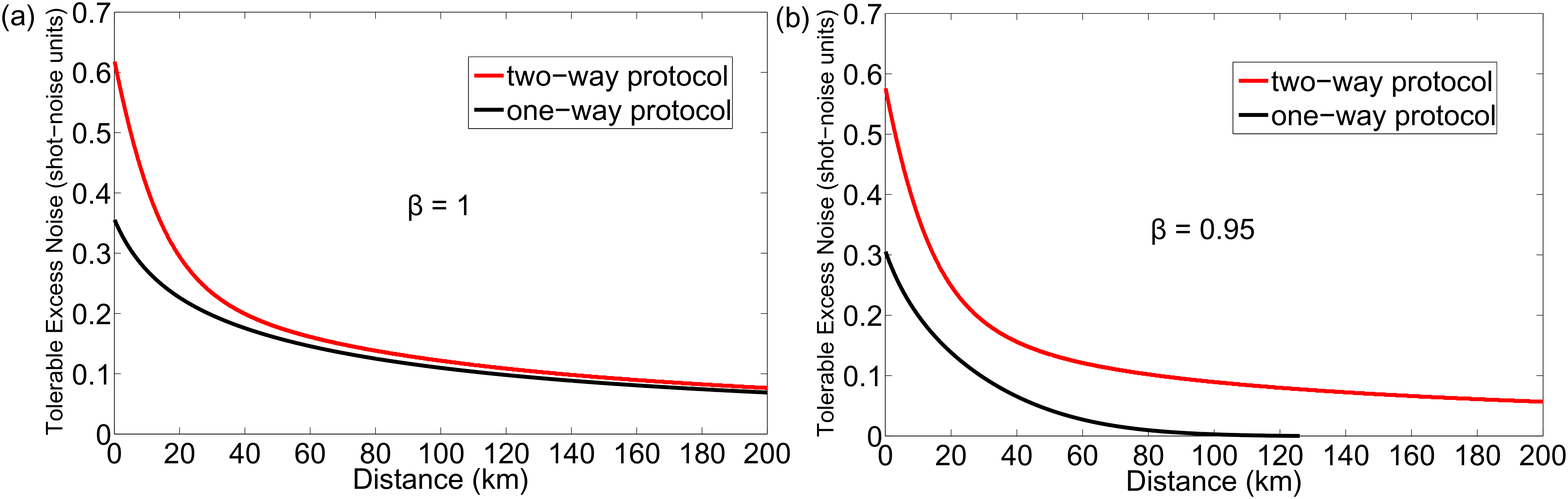}}
\caption{(Color online) (a) A comparison among the tolerable excess noise as a function of the transmission distance for two-way CV-QKD protocol against the optimal two-mode attacks and one-way CV-QKD protocol when using ideal reconciliation efficiency $\beta = 1$. (b) A comparison among the tolerable excess noise as a function of the transmission distance for two-way CV-QKD protocol against the optimal two-mode attacks and one-way CV-QKD protocol when using practical reconciliation efficiency $\beta = 0.95$~\cite{Jouguet_nature_2013}. Here we use the modulation variance $V = V_A = V_B = 20$ and $\eta = 0.75$.
}\label{fig6}
\end{figure*}

To reveal the advantage of two-way CV-QKD protocol more explicitly, the relationship between the tolerable excess noise and the transmission distance are shown in Fig.~\ref{fig6}. When using ideal reconciliation efficiency $\beta = 1$, the upper limit of the tolerable excess noise of two-way CV-QKD protocol is almost the double of the upper limit of the one-way protocol at short transmission distance. As the transmission distance increases, this advantage would be reduced. But the tolerable excess noise of the two-way protocol is still higher than that of the one-way protocol. As the reconciliation efficiency decreases from an ideal value to a more practical one~\cite{Jouguet_nature_2013}, the advantage of two-way protocols still holds and becomes more obvious. The results in Fig.~\ref{fig6} (b) show that the two-way CV-QKD could tolerate more channel excess noise than the one-way protocol.

Thus, from the above discussion, it is found that even under the optimal two-mode attacks, which cause lower secret key rate and shorter transmission distance, the performances of the two-way protocol are still better than the one-way protocol. The two-way protocol is able to distribute secret keys in communication lines which are too noisy for the one-way protocol.

\section{\label{sec:5} CONCLUSION}
In this paper, we analyze the security of two-way CV-QKD protocol against general two-mode attacks, including two independent attacks, all separable attacks and all entangled attacks. Against all accessible two-mode attacks, the expression of secret key rates of the two-way CV-QKD protocol is derived under the reverse reconciliation. Then we evaluate and compare the performance of the two-way protocol against different attacks and it is found that there is an optimal two-mode attack to minimize the performance of the protocol in terms of both key rates and maximal transmission distances. We identify the optimal two-mode attack, give the specific attack model of the optimal two-mode attack and show the performance of the two-way protocol against the optimal two-mode attack.

The performance of the two-way CV-QKD protocol against the optimal attack are compared with the performances of the one-way version of the scheme and show that the two-way CV-QKD protocols still achieve higher secret key rate and tolerate more excess noise than one-way protocol, which shows the advantage of making a double use of the quantum channel and addressing the question - whether the two-way protocol really have advantages than one-way protocol?

\section*{Acknowledgments}
This work was supported in part by the National Basic Research Program of China (973 Pro-gram) under Grant 2014CB340102, in part by the National Natural Science Foundation under Grants 61225003, 61531003, 61427813, 61401036, 61471051, in part by BUPT Excellent Ph.D. Students Foundation (CX2015205) and in part by Youth research and innovation program of BUPT(2015RC12).

\appendix
\section{Optimal two-mode attack strategy at different distance}

\begin{figure*}[!t]
\centerline{\includegraphics[width=16cm]{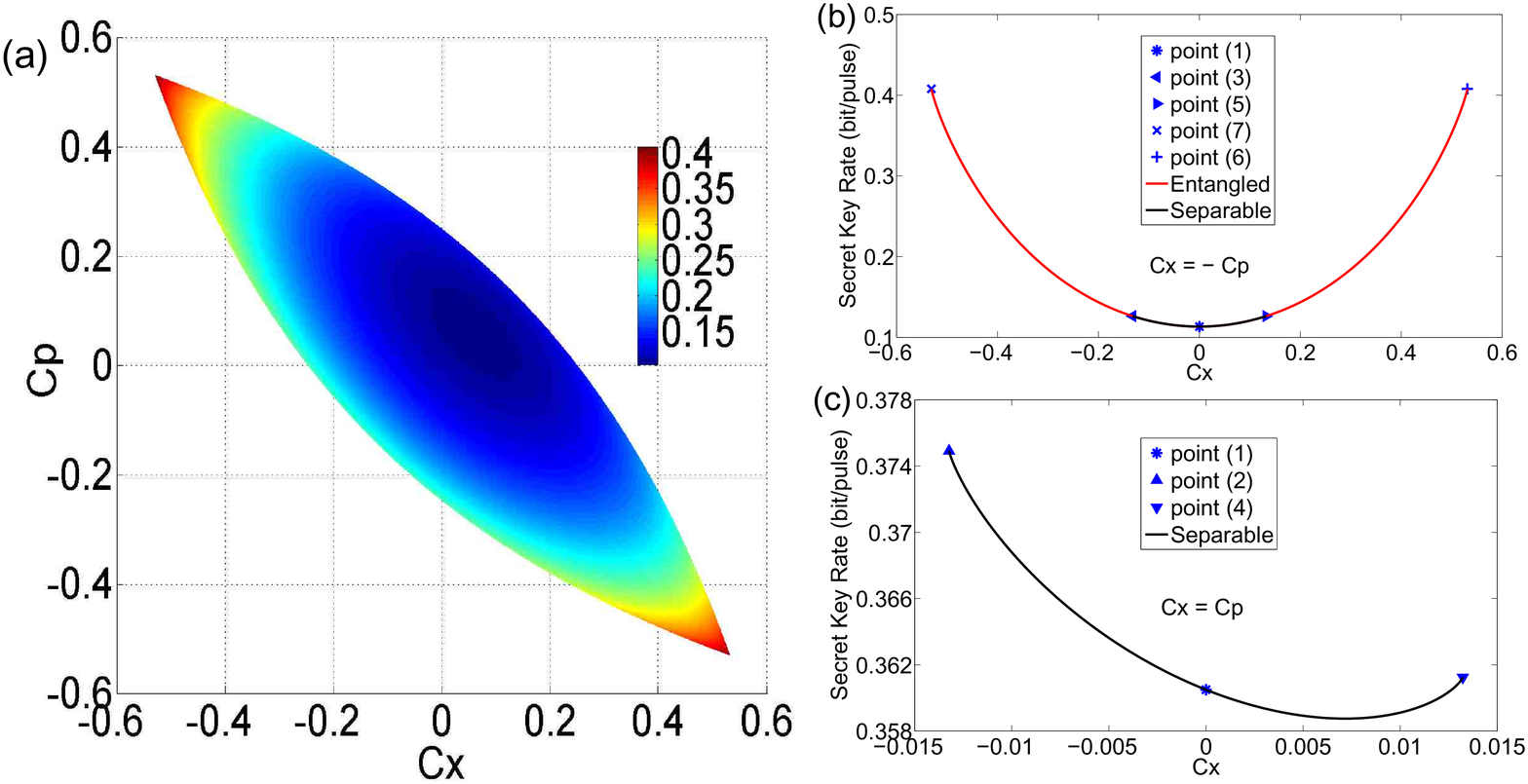}}
\caption{(Color online) (a) Secret key rates of two-way CV-QKD protocol using coherent states against all accessible attacks under $20 km$ distance. (b) Specific case of the left figure where $C_x = C_p$. (c) Specific case of the left figure where $C_x = - C_p$. Here we use the same parameters as in Fig. \ref{fig4}.
}\label{fig10}
\end{figure*}

\begin{figure*}[!b]
\centerline{\includegraphics[width=16cm]{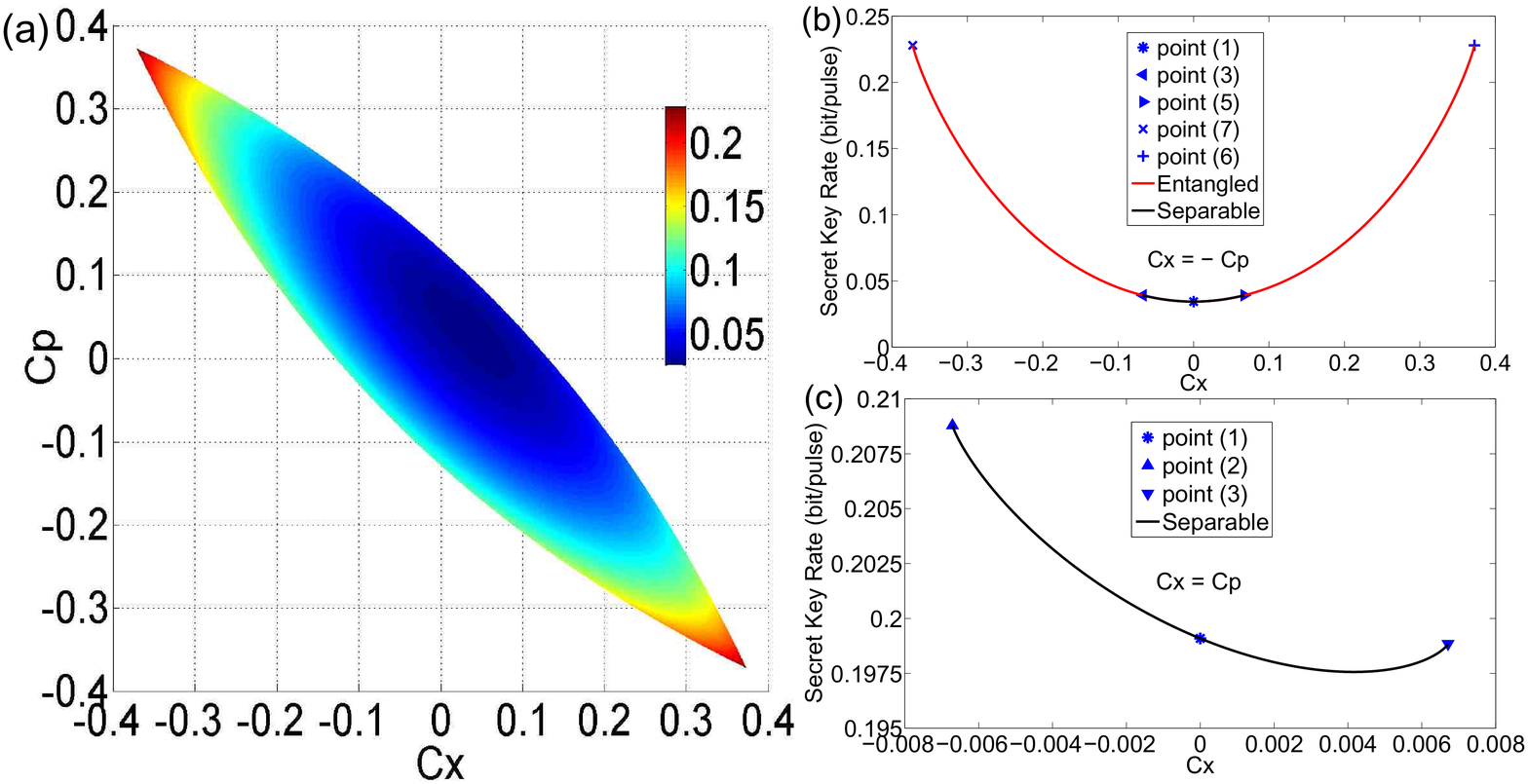}}
\caption{(Color online) (a) Secret key rates of two-way CV-QKD protocol using coherent states against all accessible attacks under $30 km$ distance. (b) Specific case of the left figure where $C_x = C_p$. (c) Specific case of the left figure where $C_x = - C_p$. Here we use the same parameters as in Fig. \ref{fig4}.
}\label{fig11}
\end{figure*}

In this appendix, we show the secret key rates of the two-way CV-QKD protocol against all accessible attacks at fixed transmission distance $d = 20 km, 30 km$. The parameters we use here keep the same with Sec. 3. As illustrated in Fig.~\ref{fig10} (a) and Fig.~\ref{fig11} (a), different colors correspond to different values of the secret key rate, where color red regions correspond to higher values of the rate, while color blue regions correspond to lower values of the rate. What's more, we also plot the specific cases where $C_x = - C_p$ and $C_x = C_p$ in Fig.~\ref{fig10} (b), (c), and Fig.~\ref{fig11} (b), (c), respectively. The minimal key rate associated with the optimal attack correspond to the two-mode symmetric separable attack $C_x = C_p = C_{opt}^{20km} = 0.0073$ at $d = 20 km$ and correspond to the two-mode symmetric separable attack $C_x = C_p = C_{opt}^{30km} = 0.0039$ at $d = 30 km$.

\newpage

\section*{References}
\begin{thebibliography}{99}


\bibitem{Gisin_RevModPhys_2002}
Gisin~N, Ribordy~G, Tittel~W and Zbinden~H 2002 \RMP \textbf{74} 145


\bibitem{Scarani_RevModPhys_2009}
Scarani~V, Bechmann-Pasquinucci~H, Cerf~N~J, Du\v{s}ek~M, L\"utkenhaus~N and Peev~M 2009 \RMP \textbf{81} 1301


\bibitem{Braunstein_RevModPhys_2005}
Braunstein~S~L and Loock~P~van, 2005 \RMP \textbf{77} 513


\bibitem{Xiang-Bin_PhysReport_2007}
Wang~X~B, Hiroshima~T, Tomita~A, Hayashi~M, 2007 Phys.~Rep.~\textbf{448} 1


\bibitem{Weedbrook_RevModPhys2012}
Weedbrook~C, Pirandola~S, Garc\'ia-Patr\'on~R, Cerf~N~J, Ralph~T~C, Shapiro~J~H and Lloyd~S 2012 \RMP \textbf{84} 621


\bibitem{Jouguet_nature_2013}
Jouguet~P, Kunz-Jacques~S, Leverrier~A, Grangier~P and Diamanti~E 2013 \emph{Nat.Photon.} \textbf{7} 378


\bibitem{Zhengyu_PhysRevA_2014}
Li~Z, Zhang~Y~C, Xu~F, Peng~X and Guo~H 2014 \emph{Phys. Rev. A} \textbf{89} 052301


\bibitem{My_PhysRevA_2014}
Zhang~Y~C, Li~Z, Yu~S, Gu~W, Peng~X and Guo~H 2014 \emph{Phys. Rev. A} \textbf{90} 052325


\bibitem{Pirandola_NatPhoton_2015}
Pirandola~S, Ottaviani~C, Spedalieri~G, Weedbrook~C, Braunstein~S~L, Lloyd~S, Gehring~T, Jacobsen~C~S and Andersen~U~L 2015 \emph{Nat.Photon.} \textbf{9} 397


\bibitem{Grosshans_PhysRevLett_2002}
Grosshans~F and Grangier~P 2002 \PRL \textbf{88} 057902


\bibitem{Weedbrook_PhysRevLett_2004}
Weedbrook~C, Lance~A~M, Bowen~W~P, Symul~T, Ralph~T~C and Lam~P~K 2004 \PRL \textbf{93} 170504


\bibitem{Navascues_PhysRevLett_2006}
Navascu\'es~M, Grosshans~F and Ac\'?n~A 2006 \PRL \textbf{97} 190502


\bibitem{Garcia_PhysRevLett_2006}
Garc\'ia-Patr\'on~R and Cerf~N~J 2006 \PRL \textbf{97} 190503


\bibitem{Pirandola_PhysRevLett_2009}
Pirandola~S, Braunstein~S~L and Lloyd~S 2009 \PRL \textbf{101} 200504


\bibitem{Renner_PhysRevLett_2009}
Renner~R and Cirac~J~I 2009 \PRL \textbf{102} 110504


\bibitem{Leverrier_PhysRevLett_2013}
Leverrier~A, Garc\'ia-Patr\'on~R, Renner~R and Cerf~N~J 2013 \PRL \textbf{110} 030502


\bibitem{Leverrier_PhysRevLett_2015}
Leverrier~A 2015 \PRL \textbf{114} 070501


\bibitem{grosshans_nature_2003}
Grosshans~F, Van Assche~G, Wenger~J, Brouri~R, Cerf~N~J and Grangier~P 2003 \emph{Nature} \textbf{421} 238


\bibitem{Lance_PRL_2005}
Lance~A~M, Symul~T, Sharma~V, Weedbrook~C, Ralph~T~C and Lam~P~K 2005 \PRL \textbf{95} 180503


\bibitem{Lodewyck_PhysRevA_2007}
Lodewyck~J, Bloch~M, Garc\'ia-Patr\'on~R, Fossier~S, Karpov~E, Diamanti~E, Debuisschert~T, Cerf~N~J, Tualle-Brouri~R, McLaughlin~S~W and Grangier~P 2007 \emph{Phys. Rev. A} \textbf{76} 042305


\bibitem{Khan_PhysRevA_2013}
Khan~I, Wittmann~C, Jain~N, Killoran~N, L\"utkenhaus~N, Marquardt~C and Leuchs~G 2013 \emph{Phys. Rev. A} \textbf{88} 010302



\bibitem{pirandola_nature_2008}
Pirandola~S, Mancini~S, Lloyd~S and Braunstein~S~L 2008 \emph{Nat. Phys.} \textbf{4} 726


\bibitem{sunmaozhu_WorldScientific_2012}
Sun~M, Peng~X, Shen~Y and Guo~H 2012 \emph{Int. J. Quantum Inform.} \textbf{10} 1250059


\bibitem{Tianyi_JPB_2014}
Wang~T, Yu~S, Zhang~Y~C, Gu~W and Guo~H 2014 \JPB \textbf{47} 215504


\bibitem{My_JPB_2014}
Zhang~Y~C, Li~Z, Weedbrook~C, Yu~S, Gu~W, Sun~M, Peng~X and Guo~H 2014 \JPB \textbf{47} 035501


\bibitem{sunmaozhu_JPB_2013}
Sun~M, Peng~X and Guo~H 2013 \JPB \textbf{46} 085501


\bibitem{Weedbrook_PhysRevA_2014}
Weedbrook~C, Ottaviani~C and Pirandola~S 2014 \emph{Phys. Rev. A} \textbf{89} 012309


\bibitem{Carlo_PhysRevA_2015}
Ottaviani~C, Mancini~S and Pirandola~S 2015 \emph{Phys. Rev. A} \textbf{92} 062323


\bibitem{Carlo_SciRep_2016}
Ottaviani~C and Pirandola~S 2016 \emph{Sci. Rep.} \textbf{6} 22225


\bibitem{Pirandola_PhysRevA_2009}
Pirandola~S, Serafini~A and Lloyd~S 2009 \emph{Phys. Rev. A} \textbf{79} 052327


\bibitem{Pirandola_OpenSystInfDyn_2013}
Spedalieri~G, Ottaviani~C and Pirandola~S 2013 \emph{Open Syst. Inf. Dyn.} \textbf{20} 1350011


\bibitem{Ottaviani_PhysRevA_2015}
Ottaviani~C, Spedalieri~G, Braunstein~S~L and Pirandola, S 2015 \emph{Phys. Rev. A} \textbf{91} 022320
Pirandola~S, Serafini~A and Lloyd~S 2009 \emph{Phys. Rev. A} \textbf{79} 052327

\bibitem{Devetak_ProcRSoc_2005}
Devetak~I and Winter~A 2005 \emph{Proc. R. Soc. London Ser. A} \textbf{461} 207


\bibitem{Nielsen_QCQI}
Nielsen~M~A and Chuang~I~L 2000 \emph{Quantum Computation and Quantum Communication} (Cambridge: Cambridge University Press).




\end{thebibliography}
\end{document}